\def \calc {{\cal {C}}}
\def \cala {{\cal {A}}}
\def \calb{{\cal {B}}}
\def \cald {{\cal {D}}}
\def \vF  {{\bf {F}}}
\def \vA1  {\bf {A}_1}

\def \vA  {\bf {A}}

\def \vr  {{\bf {r}}}

\def \vv {{\bf {v}}}

\def \vecR  {{\bf {R}}}
\def \vecV  {{\bf {V}}}

\def \hatE {{\hat {E}}}
\def \hatJ {{\hat {J}}}
\def \hatV {{\hat {V}}}
\def \fop {{f^{\rm {op}}}}
\def \mr {{<r^2>^{1/2}}}
\def \vir {{<r^2>}}
\def \si {{<v^2>^{1/2}}}
\def\ltord{\hbox{$\;\raise.4ex\hbox{$<$}\kern-.75em\lower.7ex\hbox{$\sim$}
                           \;$}}
\def\gtord{\hbox{$\;\raise.4ex\hbox{$>$}\kern-.75em\lower.7ex\hbox{$\sim$}
          \;$}}

\centerline {\bf ACCRETION OF A SATELLITE ONTO A SPHERICAL GALAXY}

\centerline {\bf II. BINARY EVOLUTION AND ORBITAL DECAY}
\vskip 20pt
\centerline {Monica COLPI}
\bigskip
\centerline {Dipartimento di Fisica, Universit\`a Degli Studi di Milano, 
Milano, Italy}

\vskip 30pt
\centerline {ABSTRACT}
\vskip 20pt
We study the dynamical evolution of a satellite (of mass $M$) orbiting
around a companion spherical galaxy.
The satellite is subject to a back-reaction force 
$\vF_{\Delta}$ resulting from
the density fluctuations  excited in the primary stellar 
system. We evaluate this force using the {\it linear response theory}
developed in Colpi and Pallavicini (1997). $\vF_{\Delta}$ is 
computed in the reference frame comoving with the primary galaxy 
and is expanded in 
multipoles. To lowest order, the force depends on 
a time integral involving a 
dynamical 4-point correlation function of the unperturbed  
background. 
The equilibrium stellar system (of mass $Nm$) is described in terms of a 
Gaussian one-particle 
distribution function. 
To capture the relevant features of the physical process determining 
the evolution of  the detached binary, we introduce in  the  
Hamiltonian the harmonic potential as 
interaction potential among 
stars. The evolution of the composite system is derived solving for a 
set of ordinary differential equations; the dynamics of the satellite
and of the stars is computed self-consistently.

We determine the conditions for tidal capture of a satellite from an 
asymptotic free 
state and give an estimate of the maximum kinetic energy above which 
encounters do not end in a merger, as a function of the mass ratio
$M/Nm.$ We find that capture always leads to final coalescence.

If the binary forms as a bound pair, stability against orbital decay is
lost if the pericentric  
distance is smaller than a critical value.
This instability is interpreted 
in terms of a {\it near resonance} condition and
establishes  when the orbital Keplerian frequency becomes comparable to the 
internal frequency $\omega$ of the stellar system.

We show that before coalescence eccentric orbits become
progressively  less eccentric: The circularization is explored as a function
of mass ratio.
The time scale of binary coalescence $\tau_b$ is 
 a sensitive function of the eccentricity $e,$ for a fixed semimajor axis
$a$ and $M/Nm$ ratio: the mismatch between $\tau_b$ at $e\sim 0\,$  and 
$\tau_b$ at $e\sim 1\,$  can be very large; typically 
 $\tau_b(e\ltord 1)\sim 6\,\omega^{-1},$ 
 and the time ratio $\tau_b(0)/\tau_b(1)\gtord 5$ 
(for $M/Nm=0.05$).
In addition we find that $\tau_b$ obeys a scaling relation with $M/Nm,$  
for circular orbits: $\tau_b\propto (M/Nm)^{-\alpha}$ with 
${\alpha}\sim 0.4$.
In grazing encounters 
$\tau_b$ is nearly independent of mass.

In a comparison with Weinberg's perturbative technique
we demonstrate quantitatively that pinning the center of mass of the galaxy
would induce a much larger torque on the satellite.

\noindent  
{\it Subject headings:} galaxies: clustering -- stars: stellar dynamics

\vskip 30pt
\centerline {I. INTRODUCTION}
\vskip 25pt

In Colpi \& Pallavicini (paper I hereafter) it was shown 
that a satellite   moving
through a nonuniform stellar background  at high speed 
experiences in addition to {\it friction}
a  force
that originates from the global {\it tidal}
deformation induced by the satellite in the spherical stellar
system during its passage.
When the massive object orbits {\it outside} the companion galaxy,
only the tidal component 
of the force affects its motion: 
In the high speed limit, this force   
acting along the instantaneous position
$\vecR$  and along the velocity vector $\vecV$ 
induces energy and angular momentum losses.

The results presented in paper I however restrict to  the case
of  shortlived encounters.
In such  flybys,  
the typical interaction timescale 
is much shorter than the dynamical time of the stellar system
and this justifies the  assumption of uniform motion 
adopted  for the satellite 
and for 
the unperturbed trajectories of the stars.
In this paper we move  a step forward extending 
our analysis to the study of the orbital  evolution 
of a binary  system composed of a  satellite  and
a spherical galaxy.
In the computation of the force, 
the periodic nature of the satellite orbit 
is  included self-consistently.  
The force on $M$, as discussed in paper I, rises  from 
the response of the stellar background to the perturbation 
induced by the satellite itself and depends on 
a  correlation tensor involving the equilibrium stellar 
dynamics.  As the time scale of the encounter
exceeds  the dynamical time of the stars bound to the galaxy, we
 account for the self-gravity of the equilibrium
 system but neglect the self-gravity of the response (see paper I for details).
The binary can become unstable to coalescence as  
energy can be exchanged between the two members
through 
a complex mechanism involving 
resonances (Lynden-bell \& Kalnajs 1972; Tremaine \& Weinberg 1984).

The evolution of a binary can proceed in  two phases:
in the early phase, the satellite orbits the companion  galaxy
while progressively losing energy; the binary is detached.
In the second more advanced phase, the satellite   
accretes onto  the stellar system: 
Moving
inside  the stellar background,  
it experiences  extensive energy losses 
by dynamical friction, spiraling 
toward the center of the galaxy.

The {\it linear response theory} (TLR hereafter) 
developed in paper I 
is an ideal tool for exploring the early phase of the dynamical
evolution of a 
satellite in the binary system. It is the aim of the paper
to determine under what  conditions
a binary loses its stability against coalescence
and how the evolution develops as a function of the mass of the satellite
and of the orbital parameters.
This analysis complements and extends an
early  work by Bontekoe and Van Albada (1987)
who explored the orbital decay
of a ``detached"  binary using a numerical
simulation.
They followed the evolution of a (softened) satellite 
moving on a close  circular orbit around 
a spherical  system  (modelled as a polytrope), and
found  that decay occurs as orbital energy is transferred into
internal energy of the primary galaxy that expands thereafter.
A numerical simulation was also performed to study  evolution in  
grazing encounters:
A satellite moving on an eccentric orbit with pericenter  internal to the
stellar system was seen to decay towards the companion after a number
of revolutions. 
Here friction intervenes during pericenter passage to cause 
decay and  progressive circularization of the orbit.

A number of problems are still unanswered
concerning the nature of the interaction between the extended
primary and the satellite.  
(a) Does the interaction always end in a merger ?
(b) When orbital stability is lost, how does evolution proceed ?
(c) How does  the decay time depend on the initial  eccentricity
and mass ratio between the satellite and
the primary ?

In the ``accretion" of a satellite onto the massive companion
the longest phase of evolution is the early phase during which
the two members are not in physical contact.
Due to the weakness of the back-reaction force
the evolution is secular, and the 
system can evolve along a sequence of quasi static states in which
the orbital parameters modify. Circularization 
preceeding infall into the system provides  an example of 
the consequences 
of the long term  evolution (Bontekoe
\& Van Albada 1987). 

These issues are of importance  in  scenarios for the formation of 
cosmic structures.
Lacey \& Cole (1997)
recognized the importance played by the values 
of the orbital parameters in affecting the evolution of baryonic cores
in merging halos.
Deriving a simplified formula for the merging time scale
in the case of a satellite moving inside a 
singular isothermal 
halo (using the Chandrasekhar expression for dynamical friction)
they showed  
that the accretion rate of baryonic  cores 
can be significantly lower
than the rate of accretion of the dark matter  halos themselves 
if satellites  fall preferentially along circular
orbits.  Thus circularization preceeding the contact phase in a binary
can affect the evolution of structures in hierarchical model of galaxy
formation (see also Navarro, Frank \& White 1994, 1995).

Cosmological N-body simulations of the rise and fall of satellites in galaxy
clusters (Tormen 1997) have also  shown that satellites  accrete onto the
primary halo preferentially  along orbits of mean circularity $\sim 0.5$.
Lighter satellites are also found to fall
on less eccentric orbits compared to the more massive ones
which  merge from nearly radial orbits.
Since the details of the ``accretion" process  can leave an imprint onto the
global shape of the cluster, this calls for a thorough analysis 
of the underlying physical mechanisms.

Numerical simulations are indeed a viable technique  for exploring the
 evolution over times comparable to a  few dynamical times of the primary.
However, accretion of light satellites 
is a secular process and spurious relaxation effect can alter the
outcome of numerical runs.
The process of orbital decay needs to be explored
using alternative methods, and 
TLR provides a framework for addressing these problems.
The method  overcomes the difficulty encountered in
previous studies (Lin \& Tremaine 1983
)
in which the galaxy was pinned to a fixed center of symmetry and applies to 
the case of a satellite moving outside the stellar distribution.
We describe the process in the frame comoving with the primary and
study the early evolution of the relative orbit.

The outline of the paper is as follows:
In $\S 2$ we compute the  equation of motion of 
 the satellite, within TLR,
expanding the force in multipoles.
We show that the force can be expressed in terms of a suitable correlation
 function.
In $\S 3$ we specify the Hamiltonian of the spherical galaxy  in its 
unperturbed state. The harmonic potential is introduced to 
describe the interaction potential of the system.
This is an idealized model in which the stellar  motion 
is characterized by a unique frequency $\omega$.
In this potential we can evaluate the back-reaction force on $M$
self-consistently, i.e.,
calculating  its value  on the 
actual dynamics of the satellite.
A binary system can form through capture of the satellite from
an asymptotic state, or can come into existence  as a bound pair.
In $\S 4$ we establish the conditions for capture 
from an initially hyperbolic orbit.
In $\S 5$ we explore the orbital evolution of a bound pair.
We show that orbital decay occurs only it the pericenter distance
is smaller than a critical value and we interpret
the result in term of a {\it near} resonant mechanism  of energy exchange
($\S 5.2$),
examining 
the evolution of circular orbits.
The secular torque for a pinned galaxy is computed
in $\S 5.3$ for a comparison with Weinberg formalism.
In $\S 6$ we then explore the dependence of the time scale of 
orbital decay 
on the initial eccentricity, showing that a large mismatch in the
scale exists between circular and very eccentric orbits. 
The dependence on the mass ratio of the binary members
is studied in $\S 6.2$
In $\S 7$ we present our 
conclusions.

\vskip 30pt

\centerline {II. BACK-REACTION FORCE}
\vskip 25pt

Consider the case
of a  satellite of mass $M$ moving in 
 the gravitational field of the primary 
system consisting
of $N$ stars of mass $m.$  
In the frame of reference comoving with the center of mass
of the stellar distribution 
the equation of motion for the satellite is 
$$\mu{d^2\vecR(t)\over dt^2}=-GMNm{\vecR(t)\over \vert \vecR(t)\vert^3}
+\vF_{\Delta}\eqno (1)$$
where $\vecR(t)$ is the position vector at time $t$ (relative to this frame)
and $\mu$ is the reduced mass $\mu=MNm/(Nm+M)$.
The force 
$$ 
\eqalignno {
\vF_{\Delta}(t)
=
& [GM]^2Nm\int_{-\infty}^t ds
\cr & 
\int d_3\vr\,d_3\vv 
\,\,\nabla_{\vv} \fop  \cdot \left [ {\vecR(s)-\vr\over
\vert \vecR(s)-\vr\vert ^3}-
{\vecR(s)\over \vert \vecR(s)\vert^3 }\right ]\,\,
{\vecR(t)-\vr(t-s)\over
\vert \vecR(t)-\vr(t-s)\vert ^3} & (2) \cr}$$
rises in response to the perturbation induced by the satellite
on the spherical system
characterized  by 
the one-particle distribution function $\fop$ 
which is isotropic and independent
of time: $\fop(\vr,\vv)$ describes the equilibrium properties of the
collisionless stellar system.
The expression (2) of the force 
is equivalent to equation (16b) of paper I modified to account
for the motion of the stellar barycenter according to equation (32); it is here 
derived,  using Gauss theorem, in the hypothesis
that the motion of the satellite is {\it external} to
the companion galaxy.

The force depends on the past history
of the satellite and  
on the response of the stellar system:
$\vr(s)$ and $\vv(s)$ denote the position and velocity
vectors of the stars at time $s$ 
as determined  by  the equilibrium Hamiltonian $H_0$. 
In equation (2) the perturbation  
vanishes as $t\to -\infty$.

Since  the satellite distance $R$ exceeds the stellar
mean radius $<r^2>^{1/2}$ we can expand the force
in  multipoles.
We thus  evaluate $\vF_{\Delta}$ expanding
in series 
the terms of the form 
$${R^a-r^a\over \vert \vecR-\vr\vert^3}
={R^a\over R^3}+
Q^{ab}r^b+{1\over 2}O^{abc}r^br^c
\eqno (3)$$
where 
$$
Q^{ab}\equiv {3R^aR^b-\delta^{ab}R^2\over
R^5}
\eqno (4)$$
and 
$$O^{abc} \equiv -{3\over R^5}(\delta^{ab}R^c+\delta^{ac}R^b+\delta^{cb}R^a)
+{15\over R^7}R^aR^bR^c.\eqno (5)$$
In equation (2) the monopole terms  in  squared brackets  
cancel out identically. In addition, because of the isotropy
of the distribution function in the velocity space, and
of the symmetry  of equation (2)
in the exchange between $\vr\to-\vr$ and $\vv\to -\vv$
yielding $\vr(t-s)\to -\vr(t-s),$
we find that the leading 
terms 
are only those coupling
a quadrupole term ($\propto Q^{ab}r^b$)
with a octupole term ($\propto O^{abc}r^br^c$).

If we include the explicit expression of the quadrupole and octupole terms in
equation (2) we find that
the force $\vF_{\Delta}$ 
 on $M$   depends simply on the dynamics of the particles (in their
unperturbed state) through correlation tensors  of the form
$$<v^ar^b\,r^c(t-s)\,r^d(t-s)>\qquad {\rm {or}}
\qquad <v^ar^b\,r^c\,r^d(t-s)>\eqno (6)$$
where the components of $\vv$ and $\vr$ are referred at current time $t,$
and $\vr(t-s)$ at time $(t-s)$. 

Because of the isotropy of the unperturbed stellar distribution
function
the tensors depend only on four scalar functions
which we introduce as follows:
$$<v^a\,r^b\,r^c(t-s)\,r^d(t-s)>=\delta^{ab}\delta^{cd}\cala(t-s)
+(\delta^{ac}\delta^{bd}+\delta^{ad}\delta^{bc})\calb(t-s)\eqno (7)$$

$$<v^a\,r^b\,r^c\,r^d(t-s)>=\delta^{ad}\delta^{cd}\calc(t-s)
+(\delta^{ab}\delta^{cd}+\delta^{ac}\delta^{bd})\cald(t-s)\eqno (8)$$
where $\delta$ is the Kronecher symbol and $<>$ denotes the mean
over the equilibrium distribution function $\fop$.

If equations (7-8) 
are inserted in (2) we find that the back-reaction force
$\vF_{\Delta}$ depends only on the correlation function $\calb$
and reads

$$F^a_{\Delta}(t)=-[GM]^2 \,N\, m^2\beta O^{abc}(t)
\int_{-\infty}^t\,ds\,\,\calb(t-s)Q^{bc}(s)\eqno (9)$$
where the tensor $O$ is evaluated at the actual position
of the satellite, i.e., at  time $t$, and $Q$ at the earlier  time $s$
depending on $R(s)$.
In deriving equation (9) we have introduced the assumption that 
$\fop$ is Gaussian in the velocity space: Accordingly,
$\nabla_{\vv}\fop=-\beta \,m\fop,$ where  
the coefficient $\beta\equiv (m\sigma^2)^{-1},$
is a function of  $\sigma$ denoting the one-dimensional stellar dispersion 
velocity (see Paper I).
We will refer to  $\vF_{\Delta}$ as back-reaction or tidal force,
hereafter. 

We  find  that 
to leading order in the multipole expansion,
the force on the satellite is expressed in terms of a time
integral coupling  the quadrupole component  at time $s$ to 
the correlation function  $\calb$ of the unperturbed background at time
$(t-s).$ 
In this paper we will  examine the physical effect that a force
satisfying equation (9) can imprint on the motion of a satellite
once we specify the nature of the
underlying system, i.e., once we specify the interaction potential
determining the properties of $\calb$. 
In paper I, we derived the force acting on $M$ 
within the impulse approximation.
Here we wish to include the self-gravity of the stars.
Because of the complexity of the mechanism, 
we consider the simplest but concrete model in which stars
interact via a harmonic potential characterized by a proper frequency 
$\omega$. We wish to gain insight into the main physical processes 
controlling the energy transfer between the satellite and the stellar
system.

\vskip 30pt
\centerline {III. HARMONIC POTENTIAL}
\vskip 25pt
We consider the response  of 
a  spherical stellar background which is characterized
by a one-particle Hamiltonian 

$$H_0={1\over 2}mv^2 + {1\over 2}m\omega r^2,\eqno (10)$$          
where $\omega$ is the internal frequency of the stars
and is independent of radius
$r$.
The corresponding one-particle distribution function, Gaussian in  
velocity space (eq. [9])  

$$\fop=\left ( {m\beta \omega \over 2 \pi}\right )^3
{{e}}^{-\beta H_0}\eqno (11)$$ is defined
so that 
$$\int d_3\vr\,d_3\vv \fop=1.\eqno (12)$$
In the harmonic potential
orbits are degenerate
since a unique frequency characterizes the motion.
This is a simplification since the mean field  potential
of a collisionless stellar system in virial equilibrium 
allows, in general, 
for a continuum 
distribution of angular frequencies.  The harmonic potential  
describes an external force field but despite this approximation we
hope to capture the relevant characteristics of
the complex physical process of binary decay.

In the harmonic potential the stars perform bound orbits around
the center of symmetry with  a random
distribution of amplitudes and phases.
The corresponding   expression for the correlation function

$$
\eqalignno {
\calb(t-s)& = <v^x\,x(t-s)\,y\,y(t-s)>
\cr & =<v^x\,x(t-s)><y\,y(t-s)> & (13) \cr}$$
simplifies considerably since  the motions along orthogonal direction are
uncorrelated in this potential  
(we denote the
components of $\vr=(x,y,z)$ and $\vv=(v^x,v^y,v^z)$, for simplicity).
Accordingly we find

$$\calb(t-s)={\sin(2\omega(t-s))\over 2(\beta m)^2\omega^3}.\eqno (14)$$

Given $\calb$ we can compute the back-reaction force [eq. (9)]: 
The motion of the satellite  is therefore 
determined by the combined effect of the 
Keplerian force   and of 
$\vF_{\Delta}.$
The system under study is simple enough that we are able to 
construct a set of ordinary differential equation 
describing the 
dynamics of the satellite.
The calculation is self-consistent and we
do not introduce any artificial or simplifying
assumption on the motion of the satellite  and on the magnitude
of its velocity relative to the stellar dispersion velocity.
Resonances that may cause secular changes
in the orbital parameters of the satellite are thus implicitly
present in the solution.

For this purpose we define the tensor 
$$I^{bc}(t)=\int_{-\infty}^{t}\,ds \,\,\calb(t-s)Q^{bc}(s)\eqno (15)$$
satisfying the equation 
$${d^2 \,I^{bc}(t)\over dt^2} = -4\omega^2 I^{bc}(t)+{1\over
(m\beta\omega)^2}
Q^{bc}(t),\eqno (16)$$
derived  from  (14). The evolution
of the satellite  is computed 
coupling equation (16) to 
equation (1) along with (9)
$$\mu{d^2 R^a\over d t^2}=-GMNm{R^a\over \vert
\vecR\vert^3}
-[GM]^2Nm^2\beta\,I^{bc}O^{abc}.\eqno (17)$$
Equations (16) and (17) form a close set
 of ordinary differential equations for $R^a$ and $I^{bc}$ 
that can be  solved after specifying the initial conditions.

Before integration, it is useful to introduce 
dimensionless variables that are defined adopting $<r^2>^{1/2}$
and $1/\omega$ as units of length and time, respectively.
According to this choice, equations (16-17)
take the form 
$${d^2 \,I^{bc}\over dt^2} = -4 I^{bc}+{1\over 9}
Q^{bc}\eqno (18)$$

$${d^2 R^a\over d t^2}=-\left (1+{M\over Nm}\right)\,\gamma_V\, {R^a\over
\vert
\vecR\vert^3}-3 {M\over Nm}
\left (1+{M\over Nm}\right)\,\gamma_V^2\,{\cal {F}}^a\eqno (19)$$
where the coefficient 
$$\gamma_V={GNm\over <r^2>^{1/2}<v^2>}\eqno (20)$$
gives the virial relation for the
spherical galaxy in its unperturbed state:
For the harmonic system, here considered, $\gamma_V=(4\pi/3)^{1/2}$.

If the keplerian motion is confined in the $(x,y)$ plane,
the components of the back-reaction force
lie in the orbital plane as well,   
and the dimensionless vector ${\cal {F}}^a$ introduced in equation (19) 
reads 
$${\cal {F}}^x=-{6\over R^5}(I^{xx}R^x+I^{xy}R^y)+{15\over R^7}R^x
\left[I^{xx}(R^x)^2+I^{yy}(R^y)^2+2I^{xy}R^xR^y\right ]
\eqno (21)$$

$${\cal {F}}^y=-{6\over R^5}(I^{yy}R^y+I^{xy}R^x)+{15\over R^7}R^y
\left[I^{xx}(R^x)^2+I^{yy}(R^y)^2+2I^{xy}R^xR^y\right ]
\eqno (22)$$
In  (18-22) all variables are regarded as dimensionless and $\vecR$ is 
evaluated at current time $t.$
The above equations are solved numerically.
For the tensor $I$
we impose, as initial condition, $I=0$ with its derivative.
At the onset of binary
evolution, the satellite is 
either set into a {\it hyperbolic} or {\it bound} keplerian orbit. 
Hyperbolic orbits are parameterized by the values of the impact parameter $b$
and by the velocity $V$. 
Elliptic orbits, at the onset of evolution,
are uniquely specified  by the  
semimajor axis $a$ (or equivalently by the binding energy per unit mass $E$ )
and by the eccentricity $e$ (or the angular momentum per unit mass  $J$).
The perturbation on a bound orbit is switch on a time $t_0=0$.
 
\vskip 30pt
\centerline {IV. CAPTURE CROSS SECTION}
\vskip 25pt

Because of the dissipative nature of the tidal force 
(paper I) 
a satellite of mass $M$ can be {\it captured} 
from an {\it asymptotic free  state}.
Equations (18-22) are then solved numerically
to determine, for a given mass ratio $M/Nm,$ 
the maximum impact parameter $b_c$ below which capture occurs,
as a function $V$, the asymptotic
velocity.
Figure 1  shows $b_c$ (in units of $<r^2>^{1/2}$) against 
$V$ (in units of $<v^2>^{1/2}=3\sigma$), 
for $M/Nm=0.05$.  
We notice that 
$b_c$ is a monotonic function of $V$; 
it increases without limit when $V\to 0,$ but declines
as $V\to \infty$ where $b_c\to 0$.
In the high speed limit, we verified that the satellite trajectory 
can be equivalently computed using the force derived in paper I (eq. [52]).

For the case under consideration
 the critical impact parameter
$b_c$ drops below $\mr$  when $V\sim 2\si$: In 
the  real interaction the satellite would then travel through  the stellar
medium  where 
{\it friction} intervenes to slow it down (see $\S 5$ and 6 of paper I): 
A merger will therefore  inevitably ensue  if the total energy loss
by friction  equals the kinetic energy at 
infinity, i.e., if 

$${[GM]^2\over V^2}\rho_0\,\mr 
\vert \Delta {\tilde {E}}(b;\epsilon)\vert 
\sim {1\over 2}\mu\,V^2, \eqno (23)$$
where 
we estimate the (dimensionless) total energy loss suffered by the satellite 
inside the stellar distribution 
$\vert \Delta {\tilde {E}}\vert$ using the expression 
derived in paper I for the case of  a homogeneous cloud
of radius $\mr$ 
(see eq. [58] and [59] for the 
energy loss, plotted in Figure 5 of paper I).
In the expression for $\Delta {\tilde {E}}$ which is a function of 
$b,$  the minimum impact parameter $\epsilon$ is set  equal to 
$\sim GM/V^2$ (since we are in the  high speed limit);
$\rho_0$ (in eq. 
[23]) is the stellar mean density 
approximated here as $\rho_0\sim 3 Nm/(4\pi \vir^{3/2}).$
Equation (23) establishes a link between 
$b$ and the asymptotic velocity
$V;$ i.e., $b=b(V)$. 

According to the   above relation, there exists a limiting 
speed  $V_{max}$ (corresponding to $b=0$), and in turn a maximum kinetic energy,
above which the encounter does not end in a merger.
When $J>0$, the largest velocity that leads to a coalescence 
is smaller than $V_{max}.$
It is useful to  introduce the dimensionless variables $\hatV=V/\si,$ 
the specific energy and angular momentum 
$$\hatE\equiv {2 E\over \mu <v^2>}=\hatV^2\eqno (24)$$
$$\hatJ \equiv {b\over L}\hatV;\eqno (25)$$
 in these units
$\epsilon/\mr \sim \gamma_V \,M/(Nm\hatV).$
Equation (23) that applies to encounters with
$b<\mr,$ provides, equivalently,  an implicit relation between 
 $\hatE$ and the specific angular momentum
$\hatJ=\hatJ(\hatE).$

Can  determine the physical parameters necessary for the
encounter to end in a final merger ?
The diagram representing the possibilities for capture of a satellite
as a function of the initial energy $\hatE$ 
and angular momentum $\hatJ$ is given in Figure 2.
The solid line gives $\hatJ(\hatE)$ resulting from equation (23)   
with $\epsilon \sim 0.05\mr$; 
 dots denote the values of
$\hatJ$ and $\hatE$ inferred from the numerical runs (as in Figure 1).
Dots describe the condition for tidal capture in  wide
 encounters for which  $b$ exceeds the virial radius $\mr$. 
 In the ($\hatE,\hatJ$) plane, the condition for
capture can be derived combining the two curves, i.e., 
joining the dots with the continuous  line, appropriate for encounters
with $b$ smaller than the virial radius.
An equivalent plot
is reported in Binney and Tremaine (1987; page 455) 
illustrating the merger conditions obtained
from N-body calculations of binary 
encounters between equal mass  galaxies with internal structure.

$\hatE_{max}\equiv \hatV_{max}$ is an increasing function of  the 
mass 
ratio $M/Nm,$ and depends on $\epsilon,$ the
``permitted size" of the satellite. 
The values of $\hatE_{max}$  
cluster between 1 and 2.5 for mass  ratios $M/Nm$ in the interval (0.01,0.1).
These are approximate estimates of the maximum kinetic energy since
they are based on the simple homogenous model of  paper I.

In exploring the orbital evolution of capture orbits we find 
 that trapping  evolves always into a  merger, i.e., 
into a state with
$R(t)\to 0$: no bound Keplerian orbits develop 
from asymptotic free states, 
as a consequence of tidal dissipation.


\vskip 30pt

\centerline {V. EVOLUTION OF BOUND ORBITS}
\vskip 25pt


In paper I, we have shown that in flybys orbital  energy is  
transferred into the
internal degrees of freedom of the galaxy.
Along a {\it bound} orbit, do tides cause dissipation ?
Or alternatively, do they only modify the gravitational field
without causing any energy loss ?


\vskip 25pt
\centerline {\it 5.1 Circularization and orbital decay}
\vskip 20pt

We here explore  the   evolution of a 
satellite moving initially  on a Keplerian 
orbit with fixed semimajor axis $a$ (expressed below in units of $\mr$)
but different  eccentricity $e$. 
Figure 3 illustrates a collection of 
orbits with $a=3$ and   mass ratio $M/Nm=0.1$.
We find that for eccentricities $e>0.5,$ 
the satellite
grazing  the outskirts of the stellar distribution
eventually suffers complete merger.
When the satellite performs, in its motion,
a number of cycles before plunge in, we find clear {\it evidence of
circularization} of the orbit (see panel (a)). 
Only at high eccentricities, the coalescence proceeds so rapidly
to prevent circularization, as shown
in panel (b).

From our numerical inspection, 
we  also find  that a {\it merger} occurs when
the pericentric distance $$a_p=a(1-e)\eqno (26)$$ 
is close to 
a critical value $$a_{p,crit}\sim 1.6\eqno (27)$$
Thus wide bound  orbits, those with  $a\gg 1$, are unstable orbits only when
$$e\gtord (1-a_{p,crit}/a).\eqno (28)$$

Condition (28)
implies the existence of a limiting distance of closest approach: 
{\it Circular orbits} are {\it stable} unless
$a\ltord 1.6.$
Panel (c) of Figure 3 depicts the 
orbital evolution of a satellite set initially on a 
nearly  circular 
orbit ($e=0.3$) at $a=3$.
The satellite maintains always at a distance from the
galaxy and tides are not efficient to extract orbital energy.
The orbit 
displays precessional motion  since the back-reaction force  causes the 
potential
to deviate from its Keplerian value (see panel (d) of Figure 3).

With decreasing mass ratio, the tidal acceleration (scaling as $M/Nm$)
weakens in magnitude. Nevertheless,
orbital decay  occurs provided 
condition (28) (derived from  the numerical analysis)
is fulfilled.
In Figure 4a and 4b,  we track the orbital decay of a satellite with
$M/Nm=0.01,\,\,a=3$ and  different eccentricities.
In this case, a larger number of orbital cycles 
 is completed before final plunge 
thus increasing the timescale of tidal infall (compare with Figure 3a and 3b).
Panel (c) and (d)  of Figure 4  
reports on the  evolution of  eccentric orbits with $a=5.$
For lighter satellites the rate of circularization per revolution is smaller;
nevertheless they perform a larger number of cycles before accreting
onto the primary. As a result, orbits with eccentricities $e\ltord 0.7$
(approximately) tend to circularize  before coalescence.

In exploring evolution of stable orbits, those 
not ending in a merger, we find that the
tidal field on $M$ induces only precessional motion.

\vskip 25pt
\centerline {\it 5.2 Tidal drag and the role of resonances}
\vskip 25pt

Can we have a deeper understanding on the mechanisms
of energy and angular momentum exchange for the system under study ?
How our findings compare with those that can be derived using  
Weinberg's perturbative approach (1986; Tremaine \& Weinberg 1984) ?

In Weinberg's  derivation of the frictional torque experienced by a satellite
orbiting within a host galaxy, the motion was constrained
to remain circular, during secular evolution. The torque was then 
evaluated considering the momentum exchange to the stars in the limit
of ${\cal {N}}\to \infty$, where $\cal {N}$ gives the number of 
orbital cycles  experienced by the satellite. 

To gain insight into the role of resonances,
in our analysis, we compute accordingly the fractional energy loss
$\delta E/E_o$ 
that the satellite would experience when constrained
to move on a arbitrary 
circular orbit  
characterized by a 
rotational frequency  $\Omega.$
This function is computed, within our model, 
 solving for the equation of energy
loss $${dE\over dt}=\vecV\cdot \vF_{\Delta}\eqno (29)$$ 
using  equations (19) (21) and (22):
The  tidal force is  evaluated  imposing circular motion on $\vecR$
and $\vecV$ (i.e., equation (18) is not included in the integration scheme).
The resulting $\delta E/E_o$ can be regarded as a function of $\Omega$
and $\cal {N}$ since time integration can be halted after 
an arbitrary number of cycles.

After completion of one period $P=2\pi/\Omega$, we find that 
the function 
$\delta
E(\Omega)/E_o$, shown in Figure 5, displays 
a sequence of dips (spread over the entire frequency interval), 
a broad  maximum about  $\Omega\sim \omega$ followed by a decline.
Dips (at which   
$\delta E/E_o=0$) 
 occur at  frequencies
$\Omega/\omega=2,2/3,1/2,2/5,1/3...$, 
independently of 
the value 
of the satellite mass and  of the radius of the 
circular orbit; 
only the magnitude of $\delta E/E_o$ depends on these parameters.

With increasing number of  cycles, dips are found to 
sweep in the frequency space  indicating that these features
are non permanent, and the extent 
of the energy loss per cycle varies with time.
This is  a consequence of the memory effect
which is intimately related to the properties
of the correlation tensor.
Remarkably, we find that 
only in the limit of $\cal {N}\to \infty$
a  {\it resonance} at $\Omega=\omega$ develops: there,  $\delta E/E_o$ 
attains its peak value, and vanishes elsewhere.
The development of the
resonance is already  evident in our numerical runs after $\sim 50$ cycles.

Due to the simplicity of the model it is possible to derive 
an analytical expression for
the rate of energy loss. When the satellite moves on a
circular path of radius $a$, energy is transferred to the stellar system
at a rate

$$\eqalignno {
 {dE\over dt}=
& 
-[GM]^2 Nm{<v^2>\over 4 a^6}{\Omega\over \omega^3}
\cr &
\left [{\sin 2(\omega-\Omega)t\over 2(\omega-\Omega)}
-{\sin 2(\omega+\Omega)t\over 2(\omega+\Omega)}
\right ].& (30)\cr}$$
If we carry out integration over 
$\cal {N}$ cycles, the function $\delta E$ vanishes
for $\Omega\neq\omega$ when the following condition is fulfilled 
$${\cal {M}}\Omega=2{\cal {N}}\omega,\eqno (31)$$ 
which is consistent with the numerical finding.
In the limit of $\cal {N}\to \infty$  
a resonance develops

$$ {dE\over dt}=-
{\pi \over 8} [GM]^2 Nm{<v^2>\over a^6}{\Omega\over \omega^3}
\left [\delta(\omega-\Omega)-\delta (\Omega+\omega)\right ].\eqno (32)$$
One can furthermore prove 
 that higher order harmonics develop
when higher order terms in the multipole
expansion of the force (eq.[2]) are included, in the limit  
${\cal {N}}\to \infty.$ Their strength however decays faster with increasing 
distance $a.$
Equation (32) shows that
after the decay of transient phenomena (related to the
way in which  the back-reaction force is turned on at $t=0$) 
the instability against  orbital decay  ensues  only if the satellite 
happens to move 
on the circular orbit having Keplerian frequency $\Omega_K$
in resonance with
the stellar system
$$\Omega_K=\omega\eqno (33)$$
Equation (33) thus  defines the critical  radius
$a_{crit}$  at which
stability is lost along a circular orbit
$$a_{crit}=[\gamma_V(1+M/Nm)]^{1/3}.
\eqno (34) $$
The value of $a_{crit}$ depends weakly
on the mass ratio; $a_{crit}=1.3$ for $M/Nm=0.01$.

We notice that  only very light satellites 
would experience 
a tidal field so weak to maintain their  motion circular 
over  many cycles and for the stellar response to develop a resonance 
(eq. [32]).
However,  
the process of orbital energy dissipation 
can not be solely interpreted as a resonance phenomenon since
massive satellites  (those with $M/Nm\gtord 0.01$) lose stability
over a broader interval of orbital frequencies 
(as illustrated in Figure 5). Heavier 
satellites  sweep fast across the  resonance
and the transient response of the galaxy guides  evolution; 
stability is lost {\it near resonance}.
The presence of dips, i.e., ``negative interferences", 
is a new feature that can affect the 
orbital evolution of the satellite in the binary, delaying the 
process of tidal drag whenever the orbital frequency sweeps through a dip
(see $\S 6$).

The transfer of angular momentum mimics that of energy. 
Confining the motion in the ($x,y$) plane,
we find an
analogous expression for the torque on the satellite which reads 

$$\eqalignno {
\tau^z=
& 
-[GM]^2 Nm{<v^2>\over 2 a^6}{1\over \omega^3}
\cr &
\left [{\sin 2(\omega-\Omega)t\over 2(\omega-\Omega)}
-{\sin 2(\omega+\Omega)t\over 2(\omega+\Omega)}
\right ]. & (35)\cr}$$

\vskip 25pt

\centerline {\it 5.3 Secular torque in a pinned galaxy }
\vskip 20pt
In this section  we first derive an expression for the secular 
torque acting on the satellite,  
using TLR.  
In a second step, we compute the same quantity
 using  Weinberg's formalism, as an independent test.

Weinberg's perturbative method (WPM hereafter) which  
adopts a factorized distribution function for the stars, 
 provides the angular momentum loss in the case
of a galaxy whose  center is nailed down.
The loss, computed in this case,
contains a contribution, difficult to disentangle,
resulting from the coordinated
displacement of the stars due to linear momentum conservation.

We 
evaluated, within TLR, the torque, denoted with $\tau_{pin},$
as a function
of the number of cycles $\cal {N}$, for a satellite 
interacting with a primary having a pinned center of mass.
In this circumstance, the back-reaction force 
is given by 
equation (16b) of paper I: 
$$ \vF_{\Delta}(t)=-[GM]^2Nm^2\beta \int_{-\infty}^t ds
\int d_3\vr\,d_3\vv 
\,\, \fop  v^a {R^a(s)-r^a\over
\vert \vecR(s)-\vr\vert ^3}
\,\,
{\vecR(t)-\vr(t-s)\over
\vert \vecR(t)-\vr(t-s)\vert ^3}.
\eqno (36)$$
To lowest order in the multipole expansion
(we retain the dipole and quadrupole terms), we compute  the torque
on the satellite.   
Assuming circular motion in the $(x,y)$ plane, we find  
$$\tau^z_{pin}=
-{[GM]^2Nm\over 2\omega a^4}
{\Omega\over 2\pi {\cal {N}}\omega}\left ({1\over (\Omega-\omega)^2}+
{1\over (\Omega+\omega)^2}\right )\left [1-\cos
\left (2\pi {\cal {N}}{\omega\over \Omega}\right )\right ].
\eqno (37)$$
From equation (37) we again clearly infer the existence of the dips 
in correspondence of which the angular momentum loss vanishes
(the energy loss mimics this behavior as well): Dips here occur
when ${\cal {N}}\omega={\cal {M}}\Omega.$
The torque is a function of ${\cal {N}}$ and in the limit of 
${\cal {N}}\to \infty$  (or $t\to \infty$)
its expression converges to a delta function:
$$\tau^z_{pin}=-{\pi \over 2}{[GM]^2Nm\over
\omega a^4}
\left [\delta (\Omega-\omega)-\delta(\Omega+\omega)\right ].\eqno (38)$$
In the limit of $t\to\infty$ we again 
infer  the existence of the
leading resonance.

We computed the secular torque 
within WPM for completeness.
In Weinberg's work a scheme is presented 
for determining, given the expression of the unperturbed 
stellar potential 
(harmonic in the case of consideration),
the ``secular  torque" experienced by the satellite, again, 
forced  to move
on a arbitrary circular orbit. 
We find that the expression of the torque calculated to lowest order
in the multipole expansion  is  
identical to equation (38).

Despite some similarities we find that the torque 
for a pinned galaxy differs in strength to the torque
for a galaxy whose barycenter is free to move. 
In the interaction between the
satellite and the galaxy with fixed center the angular momentum loss
(of order $G^2$) scales as $a^{-4}$. In a binary, instead, the
response involves higher order multipoles (from the coupling
of quadrupolar and 
octupolar terms) and has accordingly a smaller amplitude scaling as $a^{-6}$
(White 1983; Zaritsky \& White 1988; Hernquist \& Weinberg 1989).

\vskip 30pt
\centerline {VI. ORBITAL DECAY TIME}
\vskip 20pt

Numerical N-body simulations of binary mergers  customary
describe the evolution of galaxies of  comparable mass
and follow the onset of the merger
process when the two members are just {\it near contact}.
These restrictions arise since relaxation  due to the
finiteness of the system can introduce a number of spurious effects
(see also Gelato, Chernoff  \& Wasserman 1992).

Using our simplified model, we can instead explore
the process of orbital decay  of a  satellite 
in a binary  before
friction intervenes to accelerate and complete the merger process.
Indeed, it is the  time scale during the phase of detached binary
 the {\it longest} 
 scale that determines  the magnitude of the lifetime of the binary;  
$\tau_{b}$ depends on  the energy, on 
the angular momentum  and on the mass
of the satellite $M$  relative to $Nm$.

\vskip 20pt
\centerline {\it 6.1 Binary decay time versus eccentricity}
\vskip 20pt

It is  of interest to determine the characteristic time of 
binary orbital decay $\tau_b.$ 
This time scale is  a function of the initial eccentricity, 
of the semimajor axis $a,$ 
and on   $M/Nm$, the ratio determining 
the strength of the tidal field. 
For this purpose 
we follow the dynamical evolution (until $R \to 0$) 
for  a series  of orbits
having equal energy (i.e., equal semimajor axis $a$) 
but different angular 
momentum: In Figure 6 we give 
$\tau_b$ against $e$ for $a=2$ and $M/Nm=0.1$
(dots).

We find that 
the time of coalescence 
clearly diminishes  with
increasing  eccentricity. 
As $e\to 1$ the periastron distance 
$a_p$ becomes smaller than $\mr$ and the   satellite
experiences an intense  tidal force 
suffering sudden  energy
loss (below $\mr$ higher order terms in the multipole expansion become
important and are neglected here as well as the drag by
dynamical friction: both effect would speed up the process of orbital decay.
Thus $\tau_b$ provide an upper limit).

The time of coalescence  $\tau_b$ (expressed in units of the internal
dynamical time $\omega^{-1}$) varies from $\sim 35$ (for
$e=0.2$) to $\sim 6$ (for $e=0.99$) corresponding to 
a time ratio (in the two limiting cases)  of about $\sim 6$.
This estimate is in agreement with the one resulting from 
numerical N-body  simulation of sinking satellites in binaries that we 
are now performing (Mayer, Colpi \& Governato 1997).

At eccentricities  $e\sim (1-a_{p,crit}/a)$
the time  becomes exceedingly long, since the satellite 
hits the stability limit. In real systems decay would proceed
due to the energy exchange with loosely bound stars fulfilling 
condition (31).

As illustrated in Figure 6, we find, superposed  to the
clear monotonic rise  with decreasing $e$, 
the existence of abrupt increases  of $\tau_b$:
The satellite along these orbits performs many cycles 
with pericenter $a_p\gtord \mr$ 
before sudden plunge in.
 The origin  of these peaks can be attributed
to the development of transient dips in the energy
pattern  $\delta E({\cal {N}})/E_o$, delaying the infall. 
If for the same orbital parameters we diminish the extent of the tidal
force by lowering the $M/Nm$ ratio, 
the spikes in $\tau_b(e)$ appear 
over the whole  range 
of eccentricities (as shown in Figure 6 for the case with $M/Nm=0.05$ 
and $a=2$).

In Figure 7 we plot  $\tau_b$ as a function
of the initial semimajor axis $a$ for
three values of the eccentricity $e=0.1,0.3,0.5,$ and $M/Nm=0.1$.
$\tau_b$ is found to be 
an increasing function of $a$ since the tidal field displays a steep 
dependence on distance.
Circular orbits with $a$ in the interval (1.3,1.8)
decay on a time scale which is increasing exponentially when
$a\gtord  1.8$, above resonance. 

\vskip 20pt
\centerline {\it 6.2 Binary decay time  versus $M/Nm$}
\vskip 20pt
Numerical simulations explore the process of orbital decay
of spherical systems having  comparable masses.
Here, we can investigate a wider range
to determine the dependence of $\tau_b$ on the mass ratio $M/Nm$.
In Figure 8 we 
collect the results of integrations derived 
for circular orbits near resonance with 
$a$ varying in the interval $(1.3,1.7$). The time $\tau_b$  
corresponds to 
the instant at which
$R=1.$
We find that it  displays a power law dependence
on $M/Nm$ and a fit gives
$$\tau_b \propto \left ({M\over Nm}\right )^{\alpha}\eqno (39)$$
with a  slope $\alpha\sim 0.4$  nearly independent of $a.$
A similar recipe was introduced heuristically by Cole et al. (1994)
in the description of the dynamical evolution of baryonic cores
in massive dark matter haloes. 
We plot the two upper curves only above $M/Nm=0.05$ since below this value 
the tidal field
is weak 
and the two curves start displaying  the erratic behaviour.

Figure 9a shows $\tau_b$ versus $M/Nm$ for $e=0.5$ and different 
initial semimajor axis $a\ge 2$.
The lower curve gives the decay time 
for a satellite orbiting around the primary
with a pericentric distance $a_p\sim 1$. In these grazing encounters
the tidal field induces orbital decay on a time scale
of $\sim 5$ dynamical times, regardless the value of $M/Nm$.
Instead, 
in binaries with apocenter $a_p>1$
the time of binary decay  displays a dependence on $M/Nm$.
At values  $M/Nm\gtord 0.1$   the curve is monotonic
and smooth with $\alpha\sim 0.3.$ 
In Figure 9b we consider orbits with $e=0.8$;
the erratic behaviour is present whenever the initial
orbits are wide  (i.e., $a_p>1$) and similarly to the case
with $e=0.5,$ the time $\tau_b$ increases with decreasing mass ratio.

\vskip 30pt
\centerline {VII. CONCLUSIONS}
\vskip 20pt

The results presented in this paper 
are derived under the hypothesis that 
the satellite is  interacting with a 
spherical system characterized by a unique proper frequency $\omega.$ 
In exploring evolution  along  circular orbits, 
we have shown that stability is lost when the Keplerian frequency
of the relative orbit is comparable to the internal frequency of the 
stellar system; The process of energy exchange can thus be 
described in terms of 
a near resonance condition.
This is a useful concept for interpreting the origin of the 
instability 
particularly  for light satellites, those with $M/Nm\ltord 0.01$.
In this case, the energy exchange in each revolution  is exceedingly small
and the satellite performs many orbits around
the companion galaxy: The response of the stellar 
system to the {\it periodic} perturbation thus appears as a sharp 
resonance at $\Omega_K=\omega$ in the energy diagram.
For heavier satellites  energy is transferred
more effectively on a wider  interval of
orbital frequencies about $\omega$. The response of the galaxy 
to the time dependent perturbation triggers   
orbital evolution 
before the mechanisms internal to the stellar system
produce the resonance.
Thus, heavier satellites sweep fast through the unstable region of the
frequence space.
The presence of transient dips in the energy pattern
(along circular orbits)
suggests in addition that the transfer of orbital energy is a relatively
complex process  that  produces the erratic behaviour in the time scale
of binary decay. Along orbits of increasing eccentricity,
stability  is lost more rapidly than along circular orbits
and sets in when  the pericentric 
distance lies  inside a 
critical radius corresponding to the Keplerian circular orbit with
$\Omega_K\sim \omega$. This condition is equivalent to the
one  found  
by Aarseth \& Fall  (1980) in numerical N-body simulations.

In a real  spherical galaxy,  orbits are {\it
non degenerate};  
the stellar system is therefore characterized by
a spectrum of internal frequencies $N_\omega$.
The   leading resonance at $\Omega_K\sim 
\omega$ 
important for the instability
of light satellites 
would thus be 
replaced by a superposition of resonances with the effect of 
destabilizing the binary over a wider interval
of orbital  energies.  Stars 
at the periphery of the galaxy moving with 
orbital frequency  
lower than  the mean would cause secular decay from wider orbits.
The back-reaction force, in the harmonic model, is found to depend on the 
mass ratio $M/Nm$ and on the coefficient $\gamma_V$ expressing  the virial
condition of the equilibrium system. 
We expect that in a real galaxy the force will result from
the  incoherent 
superposition 
of the different monochromatic contributions:
$$F^a_{\Delta}\propto  O^{abc}
\int d\omega N_{\omega}\int ds\,  
\calb_{\omega}(t-s) Q^{bc}(s).\eqno (40)$$
Equation (9) may  thus represent  the contribution from a single frequency.
We have  demonstrated that the back-reaction force can be  
expressed in terms 
of a dynamical 4-point correlation function $\cal {B}$ of the unperturbed 
stellar system. 
As a next step we will attempt to 
compute the back-reaction force exploring the general properties of
 the correlation function
$\calb$  of 
non degenerate 
spherical systems with the aim at exploring
the dependence of  $\vF_{\Delta}$
on the spectral distribution of stars 
and at determining 
the decay time $\tau_b$ as a function
of  the orbital parameters.  We expect that the superposition of
different spectral components will erase the discontinuities present in 
$\tau_b$ that will appear as a smooth function of the orbital
parameters. 
This study is complementary to ongoing numerical investigations
(Mayer, Colpi \& Governato 1997) designed to explore the evolution
of satellites accreting onto massive dark matter haloes.
In these simulations the 
satellite is deformable and during binary evolution
mass loss by tidal stripping (neglected in our formalism; Weinberg 1996)  
can become important. The comparison will prove useful for a deeper
understanding of the process of orbital decay.

\vskip20pt
We thank G.Gyuk, M. Vietri and I. Wasserman for
stimulating discussions, and G. Gyuk for a critical reading of the manuscript.
This work was carried out with financial support from the Italian
Ministero dell' Universit\`a e della Ricerca Scientifica e Tecnologica.

\vskip 20pt
\centerline {REFERENCES}
\vskip 20pt
\def \ref {\par\noindent\hangindent=1.5truecm}

\ref Aarseth, S.J., \& Fall, S.M. 1980, ApJ, 236,43

\ref Bontekoe, Tj. R., \& van Albada, T.S. 1987, MNRAS, 224,349

\ref Cole, S., et al. 1994, MNRAS, 271, 781 

\ref Colpi, M., \& Pallavicini, A. 1997, submitted to ApJ, (paper I)

\ref Gelato, S., Chernoff, D.F., \& Wasserman, I. 1992, 384, 15

\ref Hernquist, L., \& Weinberg, M.D. 1989, MNRAS, 238, 407 

\ref Lacey, C., \& Cole, S. 1993, MNRAS, 262, 627

\ref Lin, D.N.C., \& Tremaine, S. 1983, ApJ, 264, 364

\ref Lynden-Bell, D., \& Kalnajs, A. 1972, MNRAS, 157, 1

\ref Mayer, L., Colpi, M., Governato, F. 1997, in preparation 

\ref Navarro, J.F., Frenk, C.S., \& White, S. 1994, 267, L1

\ref Navarro, J.F., Frenk, C.S., \& White, S. 1995, 275, 56

\ref Tormen, G. 1997, to appear in MNRAS

\ref Tremaine, S., \& Weinberg, M.D. 1984, MNRAS, 209, 729

\ref Weinberg, M.D. 1986, ApJ, 300, 93

\ref Weinberg, M.D. 1996, astro-ph/9607099

\ref White, S.D.M. 1983, ApJ, 274, 53

\ref Zaritsky, D., \& White, S.D.M. 1988, MNRAS, 235, 289

\vskip 30pt
\centerline {FIGURE CAPTION}
\vskip 25pt

\noindent
{\bf Figure 1}: Critical impact parameter $b_c$ 
(in units of $\mr$) below which capture occurs, against $V$ (in units
of $\si$), the satellite  asymptotic velocity of the relative 
hyperbolic orbit; the mass ratio is $M/Nm=0.05$.  

\vskip 16 pt
\noindent
{\bf Figure 2}: Angular momentum $\hatJ$ against energy $\hatE$
(in dimensionless units). {\it Dots} denote the relation derived from
integration of equations (18-22); {\it Solid} line indicates 
the relation derived using equation (23) for $M/Nm=0.05$ and $\epsilon/
\mr=0.05.$ 

\vskip 16 pt
\noindent
{\bf Figure 3}:
A collection of orbits in the plane $(x,y$); 
the  semimajor axis (in units of $\mr$) is $a=3$ at the onset of evolution, 
and $M/Nm=0.1.$   The {\it dash-dotted} circle
indicates the size of the primary galaxy.
Distances are in units of $\mr$.
Panel (a) is for an initial eccentricity $e=0.5$;
(b) is for $e=0.6$; (c) for $e=0.3$ and (d) for $e=0.4$.

\vskip 16 pt
\noindent
{\bf Figure 4} :
A collection of orbits in the plane $(x,y$) 
for a satellite with $M/Nm=0.01$;   the {\it dash-dotted} circle
indicates the size of the primary galaxy.
Distances are in units of $\mr$.
Panels (a) and (b)
refer to orbits with initial $a=3$ and $e=0.5$ and 0.6, respectively.
Panels (c) and (d) depict the  evolution of  
orbits with $a=5$ and $e=0.8$ and 0.9, respectively.
$a$ is in units of $\mr.$

\vskip 16 pt
\noindent
{\bf Figure 5}:
$\delta E/E_o$ for ${\cal {N}}=1$ as a function of  $\Omega/\omega.$

\vskip 16 pt
\noindent
{\bf Figure 6}:
Time scale of binary decay $\tau_b$
(in units of $\omega^{-1}$)  as a function of the initial
eccentricity $e$ for $a=2$.
{\it Dots} connected by a solid line
refer to $M/Nm=0.1$; {\it Squares } connected by a dash-dotted line
refer to $N/Nm=0.05$.

\vskip 16 pt
\noindent
{\bf Figure 7}:
Dimensionless time $\tau_b$ as a function of semimajor axis $a$, for  
$M/Nm=0.1.$ {\it Triangles} correspond to $e=0.5$; {\it Dots} correspond
to $e=0.3$ and {\it Squares} correspond to $e=0.1$.

\vskip 16 pt
\noindent
{\bf Figure 8}:
$\tau_b$ as a function of $M/Nm$, for orbits with $e=0$.
From bottom to top, the semimajor axis $a$ (in units of $\mr$) 
is equal to $1.3,$
$1.4,$
$1.5,$
$1.6,$
$1.7.$

\vskip 16pt
\noindent
{\bf Figure 9}:
(9a) $\tau_b$ as a function of $M/Nm$, for orbits with $e=0.5$.
From bottom to top, the semimajor axis $a$ (in units of $\mr$) is equal to $2,$
$2.3,$
$2.5,$
$2.7.$; (9b) $\tau_b$ as a function of $M/Nm$, for orbits with $e=0.8$.
From bottom to top, the semimajor axis $a$ is equal to $5,$
$5.3,$
$5.5.$

\bye
In a real encounter
the satellite has internal structure and the action of the tidal field
of the companion galaxy can be important in inducing mass loss:
this would enhance the probability of final meager. Thus, our
results are indicative and even at eccentricities
smaller than those inferred in the calculation, capture can occur.

+++++++++++++++++++++
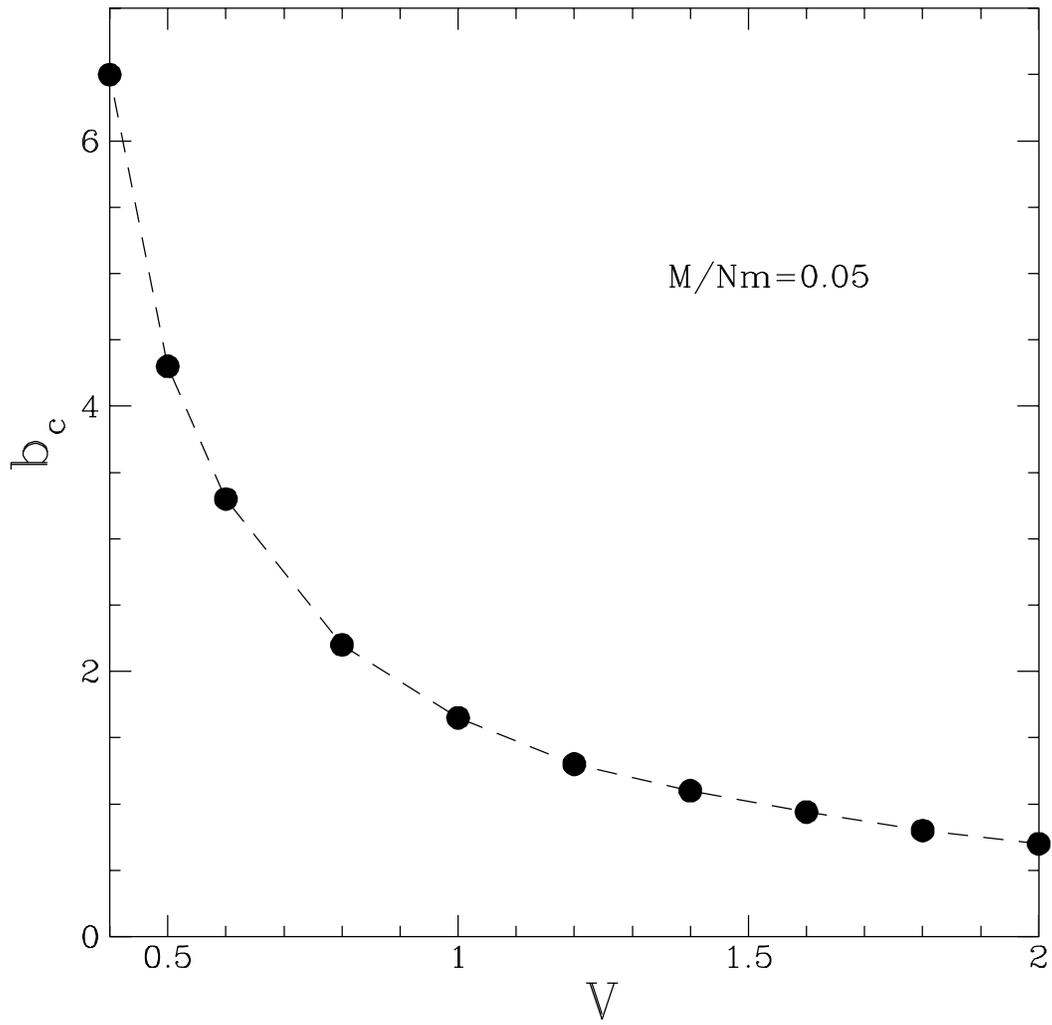
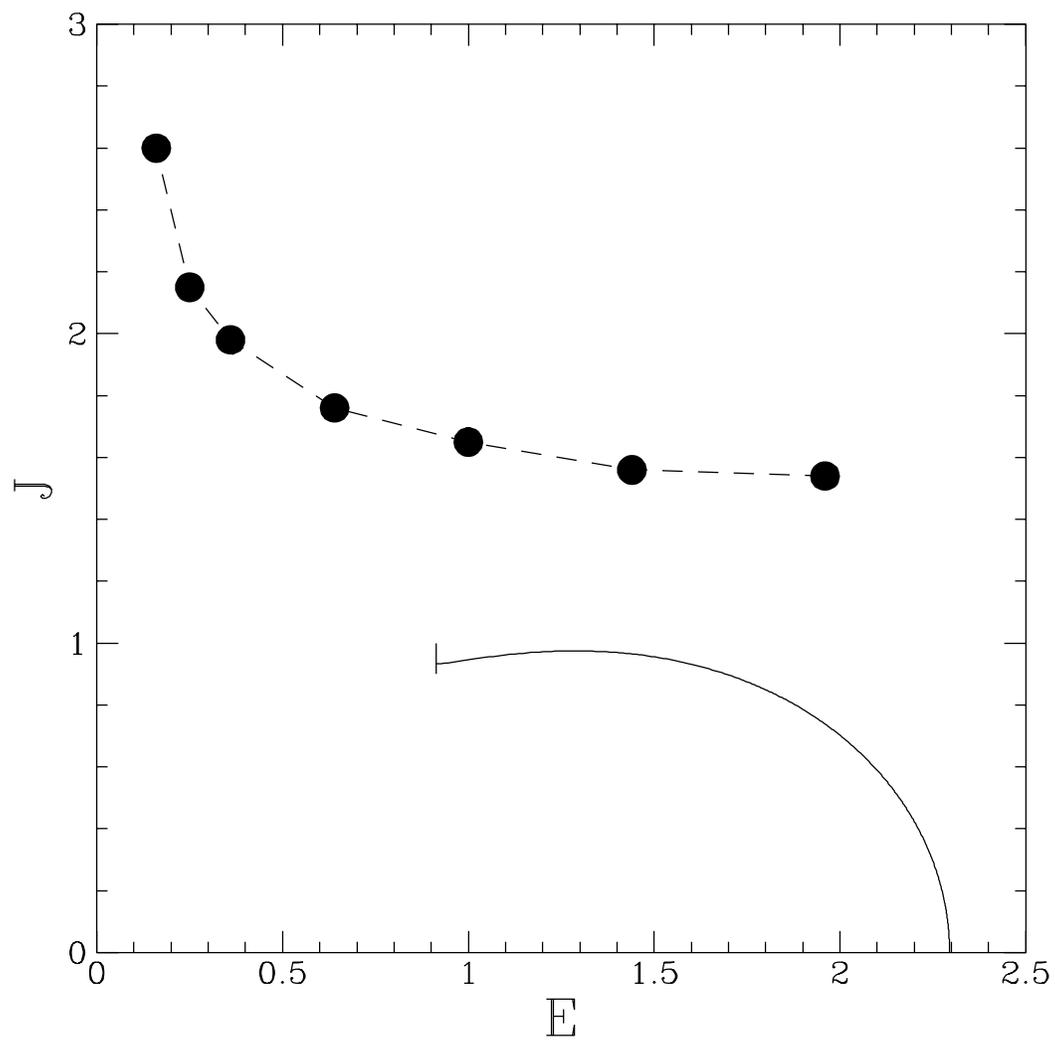
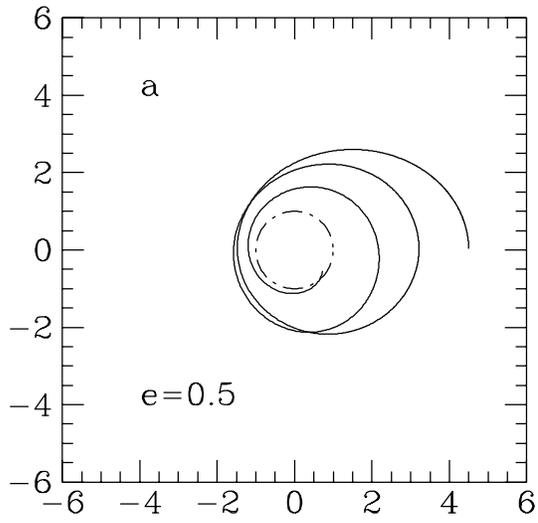
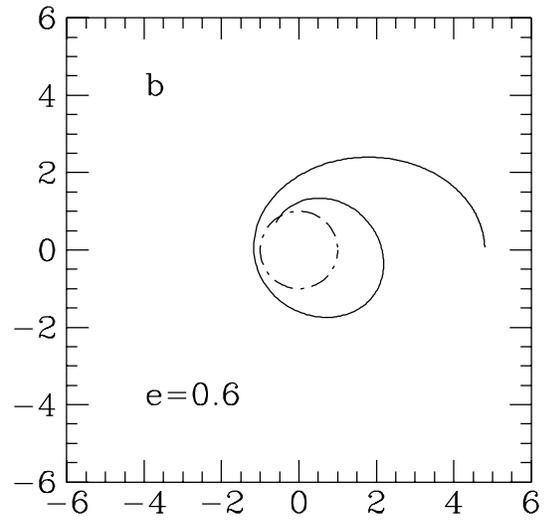
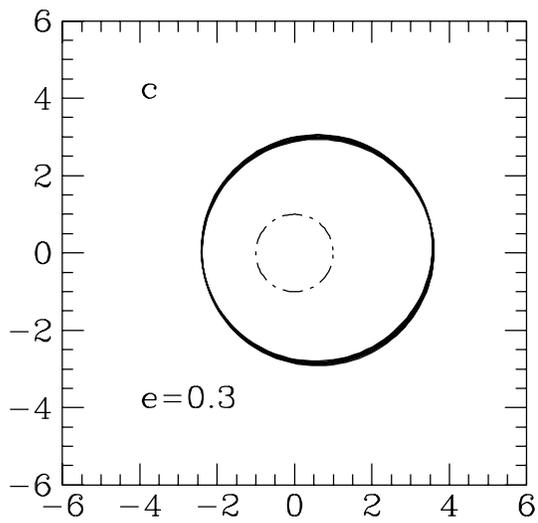
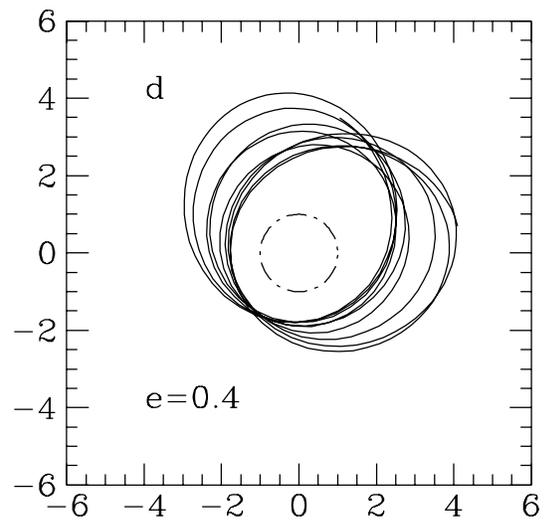
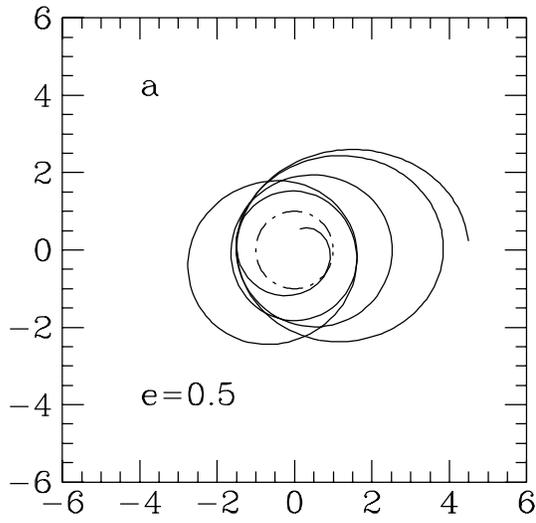
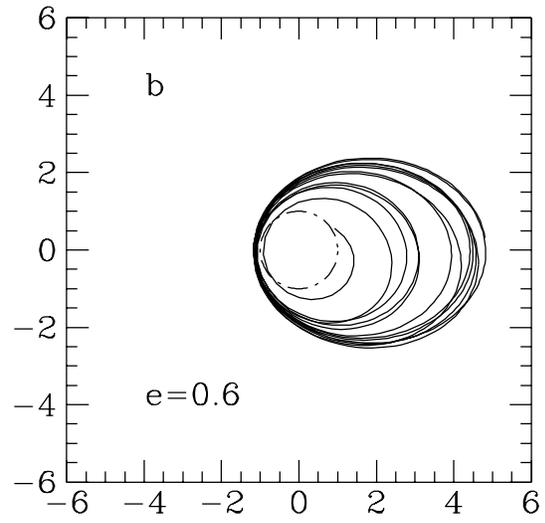
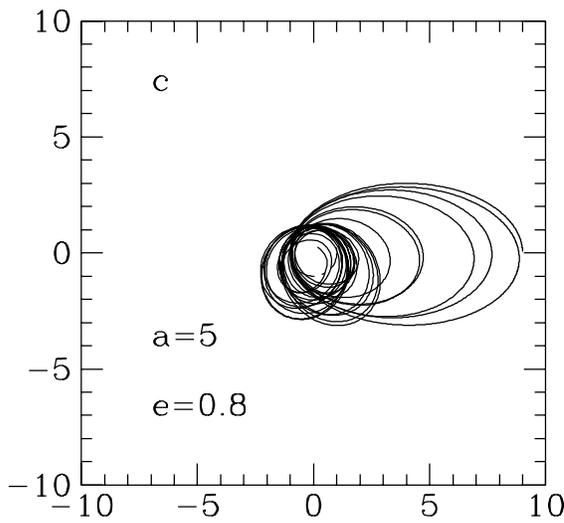
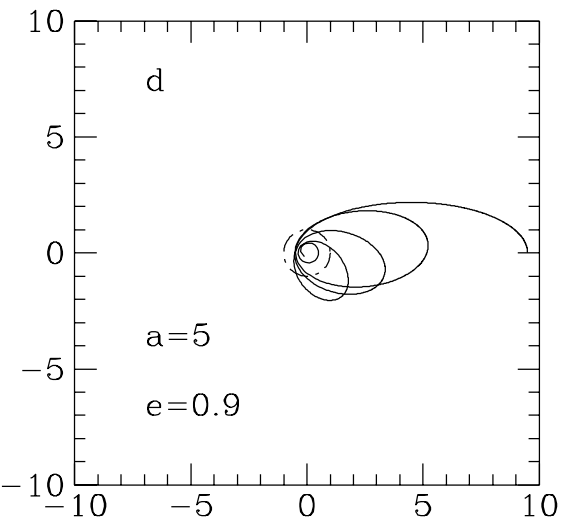
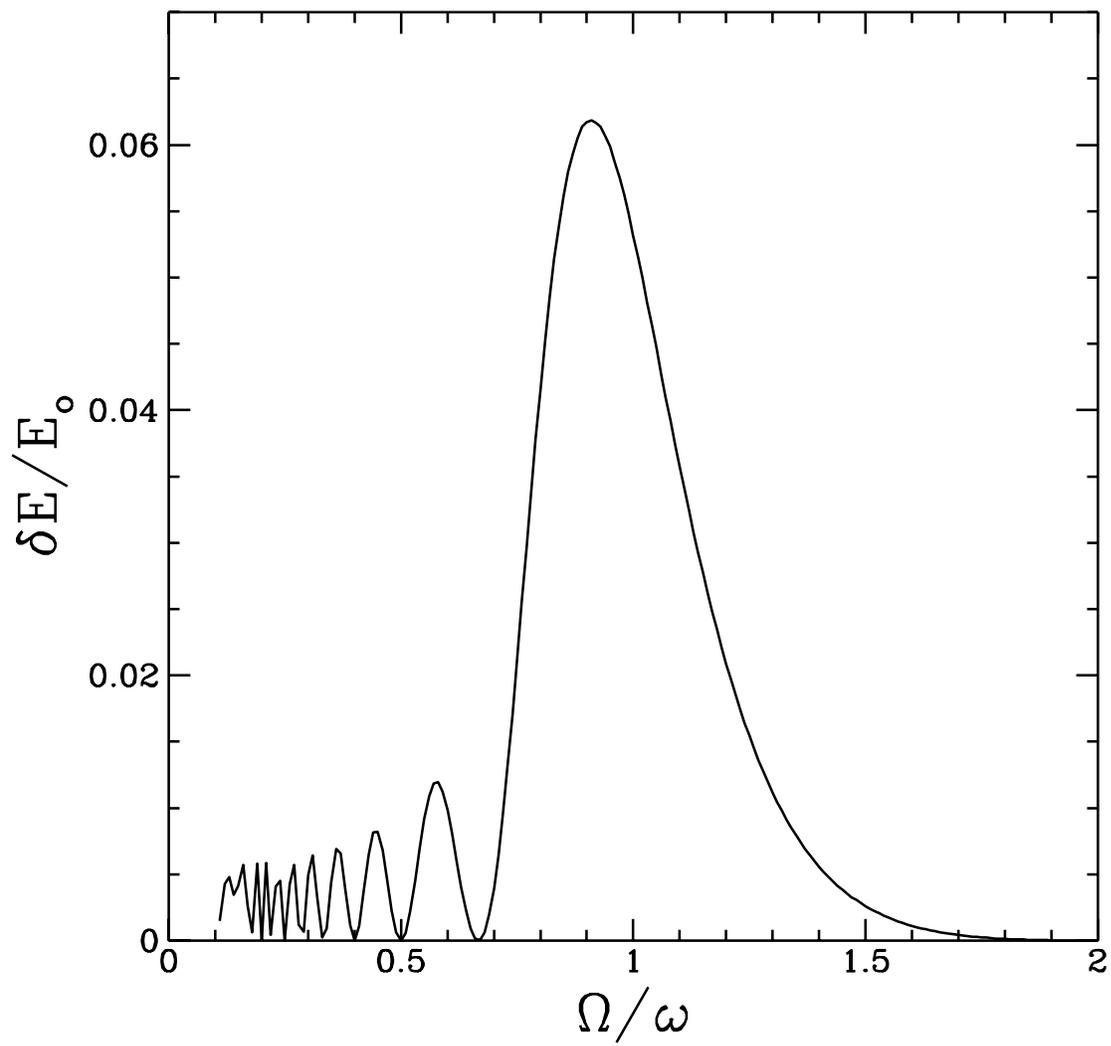
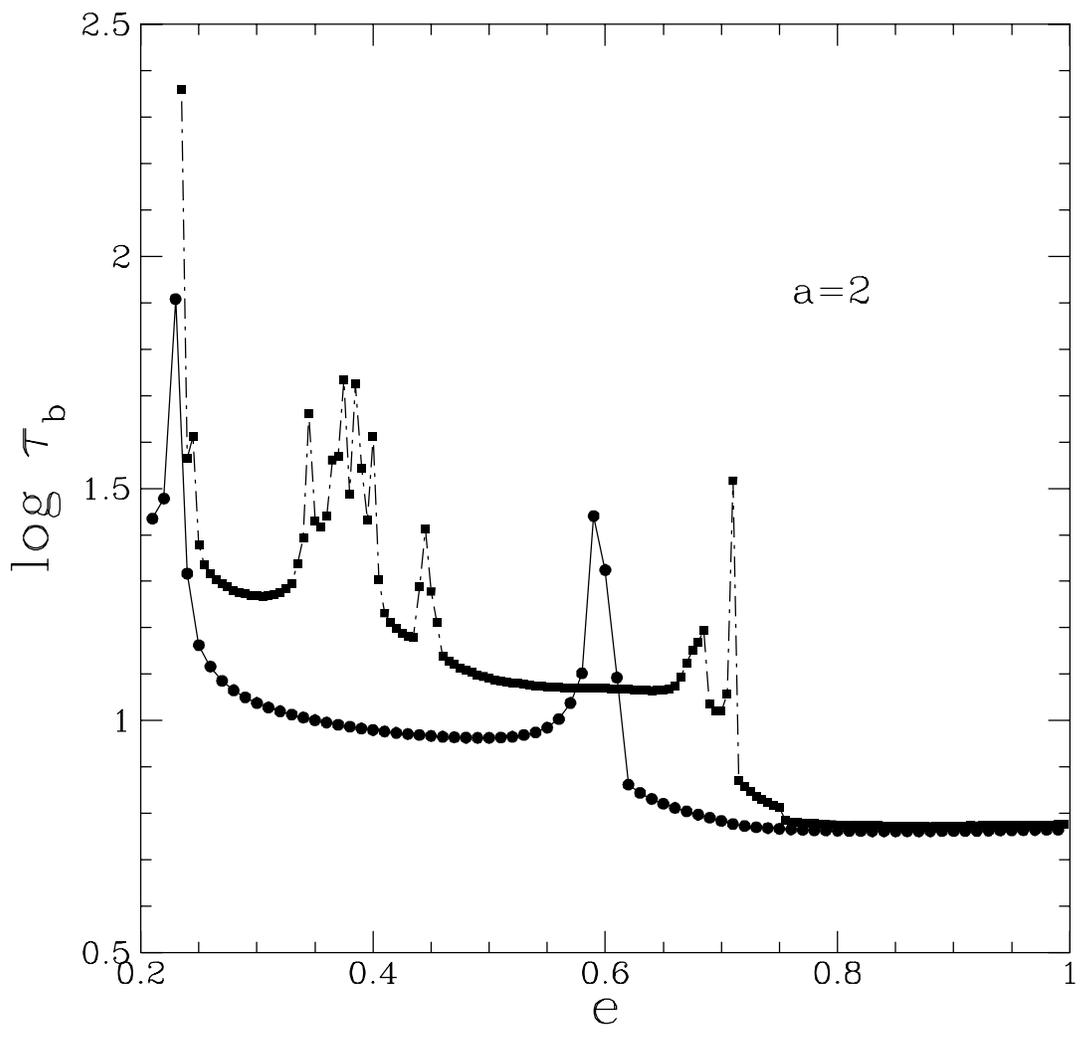
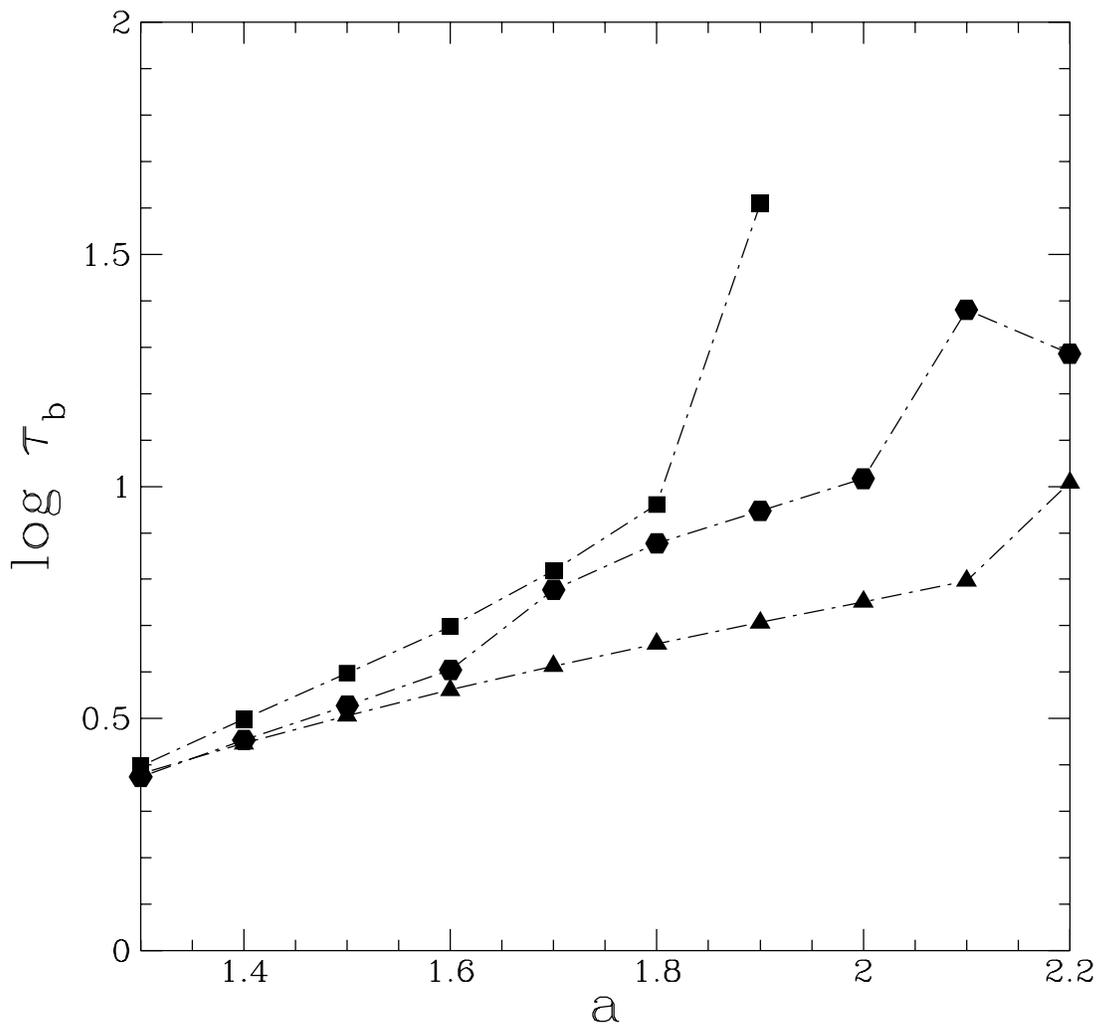
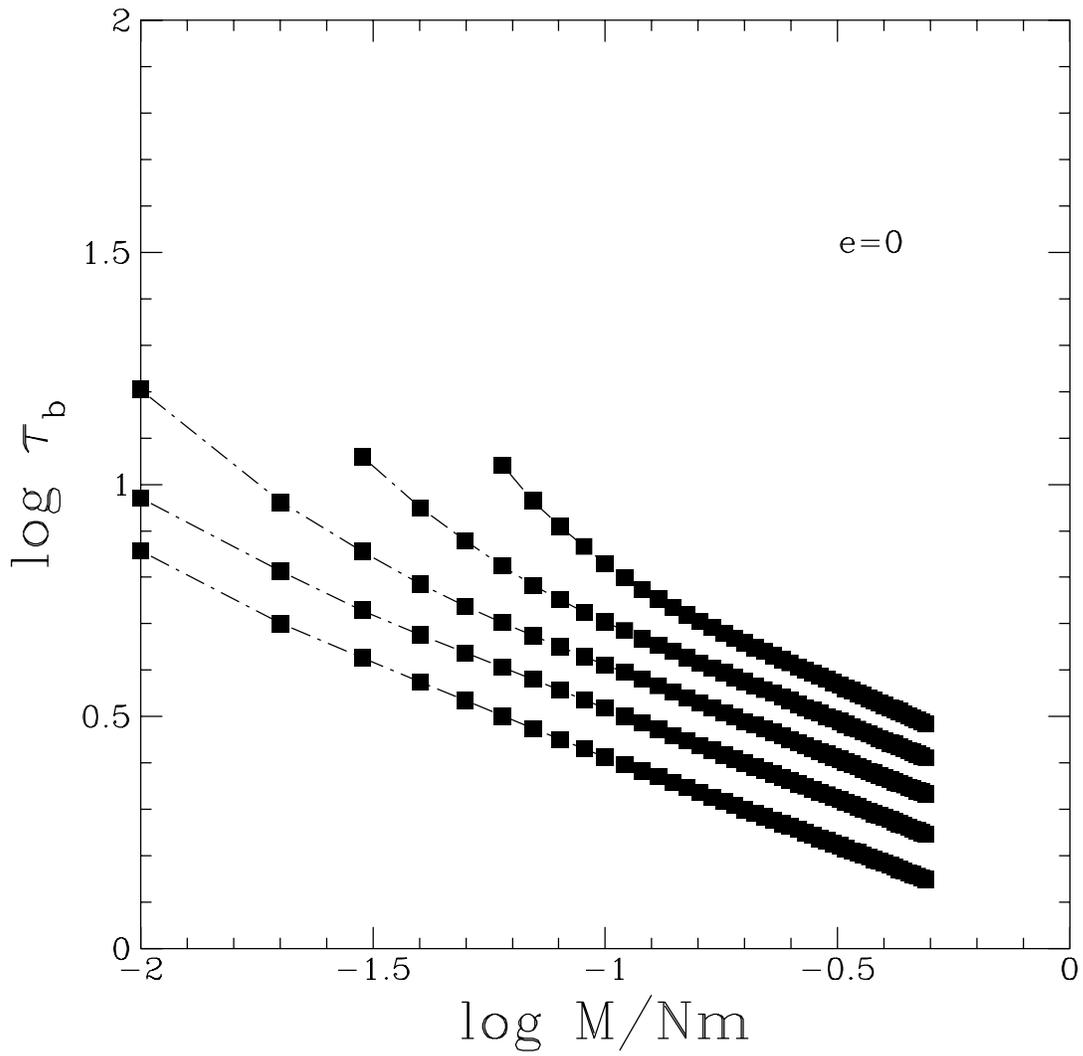
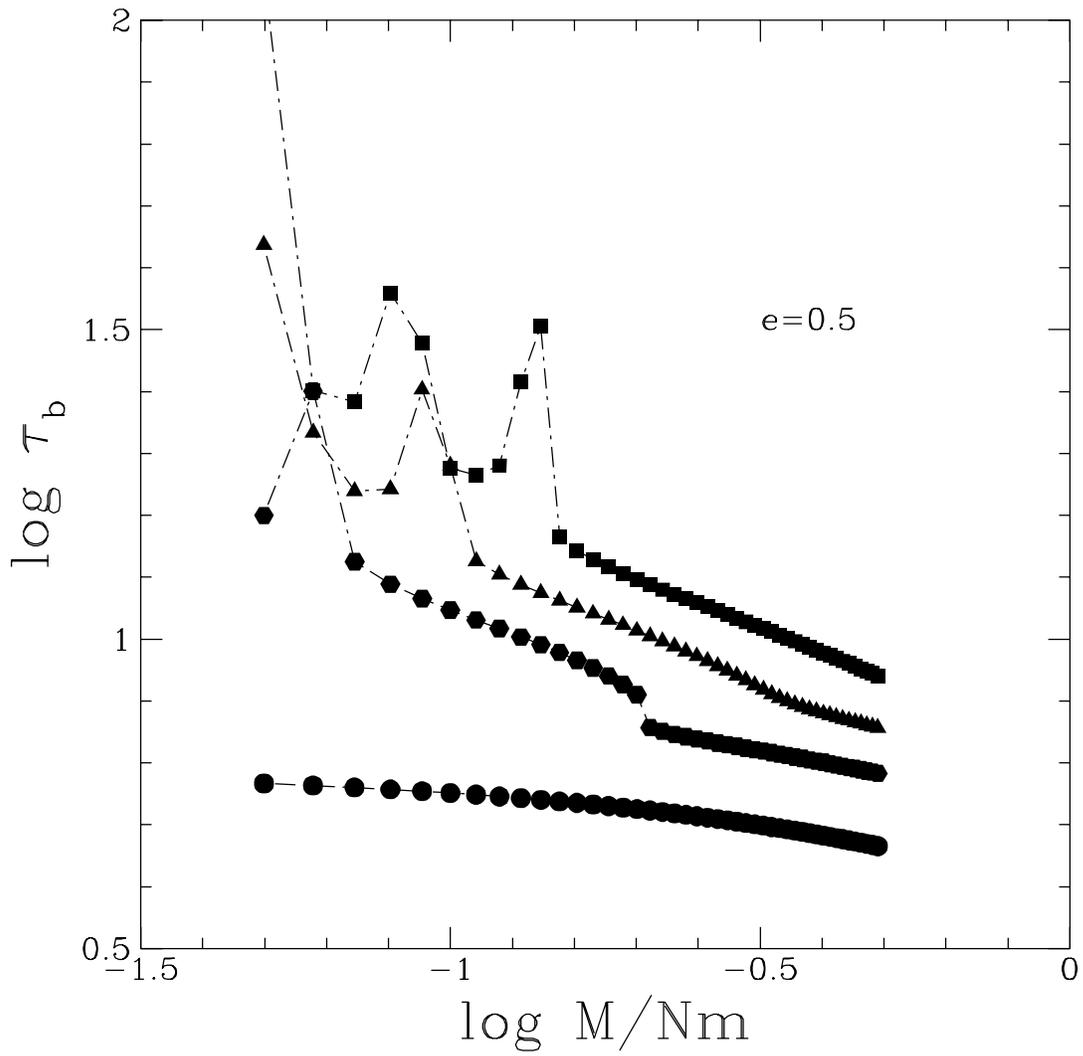
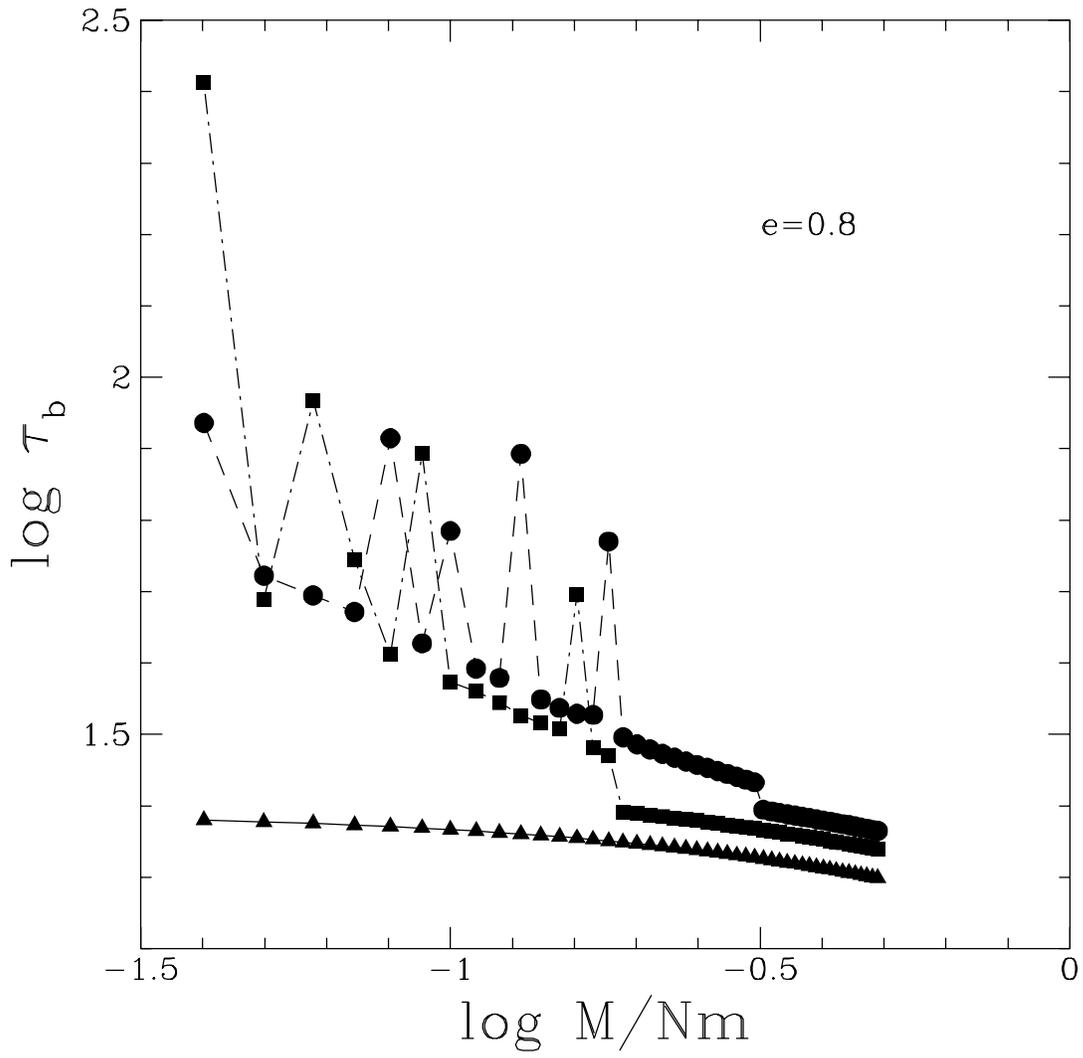
\bye